\documentclass[aps,prl,groupedaddress,superscriptaddress,twocolumn,showpacs]{revtex4-1}

\usepackage{graphicx}
\usepackage{dcolumn}
\usepackage{bm}
\usepackage[T1]{fontenc}
\usepackage[french]{babel}
\usepackage{epstopdf}
\usepackage{amssymb}
\usepackage{color}
\usepackage{amsmath}
\usepackage{subfigure}
\usepackage{natbib}

\begin{document}

\preprint{APS/123-QED}

\title{Controlling the topological sector of magnetic solitons in exfoliated Cr$_{1/3}$NbS$_2$ crystals}

\author{L. Wang}
\affiliation{Department of Quantum Matter Physics and Group of Applied Physics, University of Geneva, 24 quai Ernest-Ansermet, CH-1211 Geneva, Switzerland}

\author{N. Chepiga}
\affiliation{Institute of Physics, \'{E}cole Polytechnique F\'{e}d\'{e}rale de Lausanne, 1015 Lausanne, Switzerland}

\author{D.-K. Ki}
\affiliation{Department of Quantum Matter Physics and Group of Applied Physics, University of Geneva, 24 quai Ernest-Ansermet, CH-1211 Geneva, Switzerland}

\author{L. Li}
\affiliation{Department of Materials Science and Engineering, University of Tennessee, Knoxville, Tennessee 37996, USA}

\author{F. Li}
\author{W. Zhu}
\affiliation{Theoretical Division, T-4 and CNLS, Los Alamos National Laboratory, Los Alamos, New Mexico 87545, USA}

\author{Y. Kato}
\affiliation{Department of Applied Physics, University of Tokyo, Hongo, 7-3-1, Bunkyo, Tokyo 113-8656, Japan}

\author{O.S. Ovchinnikova}
\affiliation{Nanofabrication Research Laboraotry, Center for Nanophase Materials Sciences, Oak Ridge National Laboratory,
Oak Ridge, TN 37831-6493, USA}

\author{F. Mila}
\affiliation{Institute of Physics, \'{E}cole Polytechnique F\'{e}d\'{e}rale de Lausanne, 1015 Lausanne, Switzerland}

\author{I. Martin}
\affiliation{Materials Science Division, Argonne National Laboratory, Argonne, Illinois 60439, USA}

\author{D. Mandrus}
\affiliation{Department of Materials Science and Engineering, University of Tennessee, Knoxville, Tennessee 37996, USA}
\affiliation{Materials Science and Technology Division, Oak Ridge National Laboratory, Oak Ridge, Tennessee 37831, USA}
\affiliation{Department of Physics and Astronomy, University of Tennessee, Knoxville, Tennessee 37996, USA}

\author{A. F. Morpurgo}
\affiliation{Department of Quantum Matter Physics and Group of Applied Physics, University of Geneva, 24 quai Ernest-Ansermet, CH-1211 Geneva, Switzerland}
\email[]{Alberto.Morpurgo@unige.ch}


\begin{abstract}
We investigate manifestations of topological order in monoaxial helimagnet Cr$_{1/3}$NbS$_2$ by performing transport measurements on ultra-thin crystals. Upon sweeping the magnetic field  perpendicularly to the helical axis, crystals  thicker than one helix pitch (48 nm) but much thinner than the magnetic domain size ($\sim$1 $\mu$m) are found to exhibit sharp and hysteretic resistance jumps. We show that these phenomena originate from transitions between  topological sectors with different number of magnetic solitons. This is confirmed by measurements on crystals thinner than 48 nm --in which the topological sector cannot change-- that do not exhibit any jump or hysteresis. Our results show the ability to deterministically control the topological sector of finite-size Cr$_{1/3}$NbS$_2$ and to detect inter-sector transitions by transport measurements.
\end{abstract}

\maketitle

The properties of many electronic systems are characterized by topological indices that allow all possible states to be grouped into distinct topological sectors~\cite{Monastyrski1993,Volovik2003,Altland2010,Fradkin2013,Qi2011}. Since topological indices assume discrete values, changes in topological sector can only occur through abrupt transitions that can be detected experimentally. Investigating these transitions under controlled conditions and probing properties associated to "topological order", however, is not simple, as it requires the ability to tune the state of the system in a predefined topological sector, \textit{i.e.} to deterministically set the topological indices of the system by acting on experimental parameters. Here we show that this level of control can be achieved by means of magneto-resistance measurements on mechanically exfoliated crystals of monoaxial chiral helimagnet Cr$_{1/3}$NbS$_2$.

Our experiments build on earlier work that has established the key properties of Cr$_{1/3}$NbS$_2$, a layered material consisting of alternating NbS$_2$ and Cr planes (see Fig. 1(a))~\cite{Parkin1980,Parkin1980a,Moriya1982,Miyadai1983,Yurii1984,Kousaka2009,Togawa2012,Ghimire2013,Togawa2013,Chapman2014,Bornstein2015,Tsuruta2016}. At low temperature, the $S=3/2$ spins on the Cr atoms order ferromagnetically in each plane, forming a helix that winds around the direction perpendicular to the planes (Fig. 1(b), the zero-field helix pitch $L_0$ is 48 nm)~\cite{Miyadai1983,Kousaka2009,Togawa2012,Ghimire2013,Togawa2013}. Lorentz microscopy has shown that in-plane magnetic field $B$ causes the helix to deform, resulting in a so-called chiral soliton lattice~\cite{Togawa2012}. The lattice (Fig. 1(b)) consists of narrow regions in which the spins make a complete 2$\pi$ revolution (the solitons), separated by stretches of ferromagnetically aligned spins, whose extension --which determines the lattice period $L_B$-- increases upon increasing $B$ (Fig. 1(c))~\cite{Togawa2012}. The observed microscopic evolution of the helix, as well as the detailed magnetic response of bulk crystals, conform quantitatively to a theoretical (one-dimensional) model proposed by Dzyaloshinskii over 50 years ago, comprising Heisenberg and Dzyaloshinskii-Moriya interactions between nearest neighboring spins (besides the Zeeman term in the presence of a magnetic field and a magnetic anisotropy term that forces the spins to point perpendicularly to the helix direction)~\cite{Dzyaloshinskii1958,Moriya1960}.

\begin{figure}[t]
\includegraphics[width=8cm]{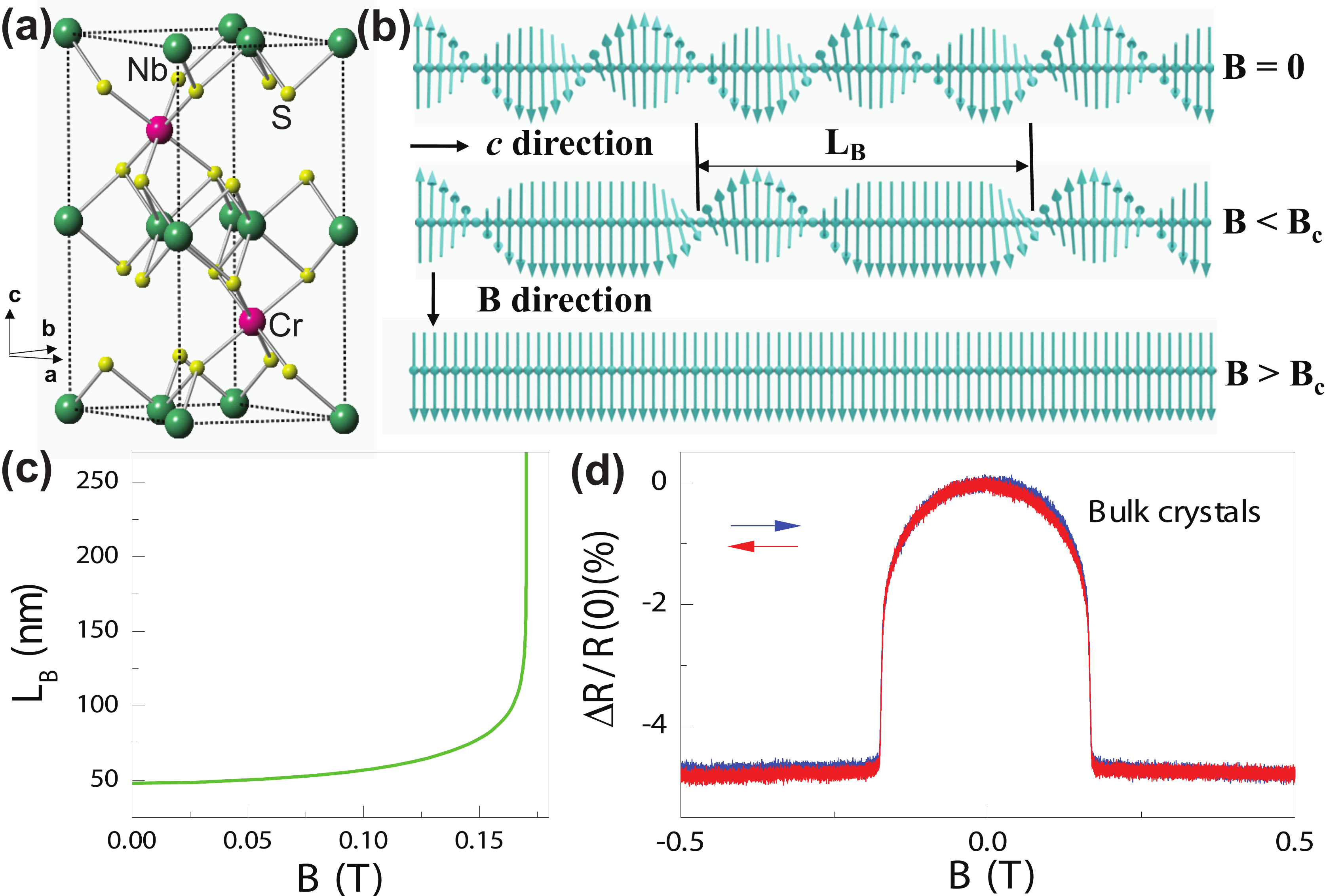}
\caption{\label{fig1} (Color online) (a) Structure of layered Cr$_{1/3}$NbS$_2$. The $c$-axis lattice constant is 1.21 nm. (b) Schematic illustration of the spin configurations along the c-axis in bulk Cr$_{1/3}$NbS$_2$ (the arrows represent the magnetization in the Cr planes). Upon increasing magnetic field ($B$) in the $ab$-plane, the $B=0$ magnetic helix  (top) is gradually transformed into a chiral soliton lattice with period $L_B$ (middle), finally becoming ferromagnetic above the critical magnetic field ($B_c$; bottom). (c) In the process, the period $L_B$ increases and diverges close to $B_c$ (plot obtained from theoretical calculations based on the model discussed in the main text). (d) Magneto-resistance, $\frac{\Delta R}{R(0)}\equiv\frac{R(B)-R(0)}{R(0)}$, of bulk  Cr$_{1/3}$NbS$_2$  measured at 250 mK. The blue/red curves correspond to measurements performed upon sweeping the in-plane field $B$ in the directions indicated by the blue/red arrows. An abrupt resistance drop at $B_c\sim0.17$ T with no hysteresis is observed.}
\end{figure}

\begin{figure}[t]
\includegraphics[width=8.5cm]{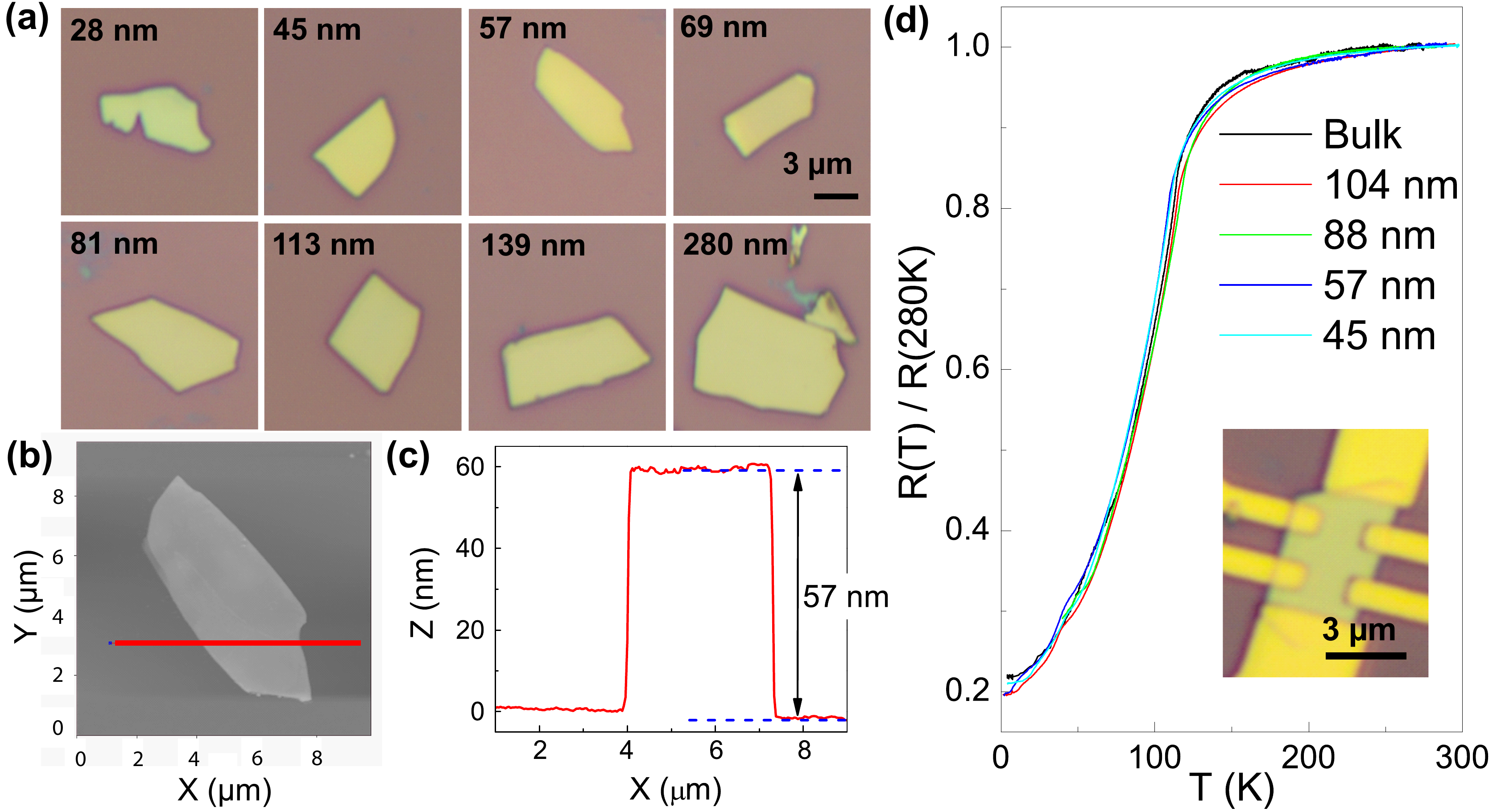}
\caption{\label{fig2} (Color online) (a) Optical microscope images of exfoliated Cr$_{1/3}$NbS$_2$ crystals with thickness between 28 nm and 280 nm. (b) Atomic force microscope image of the 57-nm-thick crystal, showing a flat and uniform surface, as visible in the thickness profile (c) taken along the red line. (d) Temperature dependence of the resistance $R(T)/R(280 K)$ for crystals of different thickness (see legend). All curves merge together and exhibit a resistance drop starting around 130 K, the bulk transition temperature to the helimagnetic state. The inset in (d) shows an optical image of a device, with nano-fabricated Ti/Au contacts. In all images, the scale bar is 3 $\mu$m.}
\end{figure}

Whenever it is spatially confined, the chiral soliton lattice in Cr$_{1/3}$NbS$_2$ is predicted to exhibit interesting phenomena~\cite{Kishine2014,Tsuruta2015,Togawa2015,Tsuruta2016a}. Some of these phenomena have been observed recently in small specimen (10 $\mu$m in linear dimensions) cut from bulk crystals, in which confinement originates from the presence of magnetic domains extending for approximately 1 $\mu$m in the helix direction~\cite{Togawa2015}. Upon increasing $B$, the separation between solitons increases, so that their total number in each finite-size domain decreases. Since the number of solitons corresponds to the total spin winding number that is a topological index, the soliton number can only change through discrete transitions. These transitions have been detected by analyzing Lorentz microscope images~\cite{Togawa2015}. It was argued that hysteresis and a sequence of jumps present in the magneto-resistance of small specimen, but absent in bulk crystals (see Fig. 1(d)), are a transport manifestation of the changes in soliton number~\cite{Togawa2015,Tsuruta2016a}. This conclusion is very interesting, as it implies the ability to probe topological aspects of the magnetic state of Cr$_{1/3}$NbS$_2$ by monitoring transport properties. Its validity is however unclear because the magneto-resistance jumps could not be directly linked to specific changes in the soliton configuration, and because it was not ruled out that the jumps may originate from domain switching.

To avoid these problems, we investigate these same phenomena by working with Cr$_{1/3}$NbS$_2$ crystals much thinner than the magnetic domain size. Fig. 2(a) shows optical microscope images of a selection of crystals having thickness ($t$) between 28 and 300 nm, whose surfaces are parallel to the NbS$_2$ and the Cr planes. The magnetic helix is oriented perpendicular to the substrate, so that the crystal thickness determines the number of solitons present at $B=0$. We produced these crystals, up to 500 times thinner than the specimen studied in Ref.~\cite{Togawa2015}, by means of mechanical exfoliation, following the same procedure used to extract graphene from graphite. Exfoliation is  more difficult for Cr$_{1/3}$NbS$_2$ because of strong chemical bonds between the Cr and S atoms; nevertheless, atomic force microscope images (see Figs. 2(b-c)) show that the crystal surface is flat and uniform.

Transport through Cr$_{1/3}$NbS$_2$ crystals as thin as the ones discussed here has not been investigated earlier and it is important to identify which properties depend on thickness and which do not. Fig. 2(d) shows that all exfoliated crystals exhibit the same temperature dependence of the resistance (identical to that of the bulk). A pronounced decrease in resistance starts around $T=130$ K, corresponding to the transition temperature $T_c$ to the helimagnetic state. Since $T_c$ is determined by the strength of the microscopic interactions between nearest neighboring spins (and --in the range investigated here-- not by the thickness) this behavior is not surprising~\cite{Bak1980}. Nevertheless, the excellent reproducibility upon varying the thickness is worth commenting, as it indicates the absence of any significant material degradation (not warranted \textit{a priori} for thin crystals exposed to air during exfoliation and device fabrication).

\begin{figure}[t]
\includegraphics[width=8.5cm]{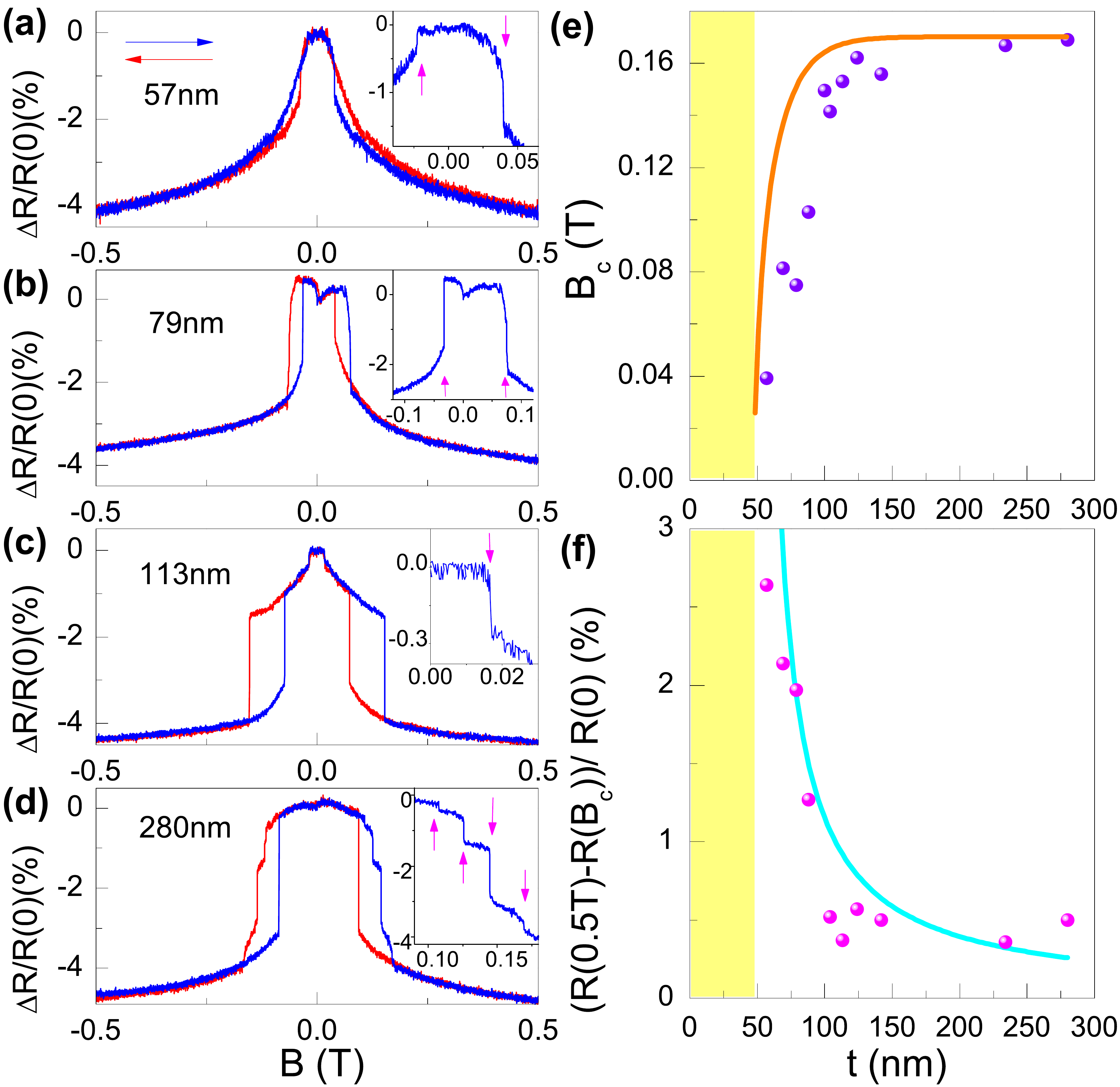}
\caption{\label{fig3} (Color online)  (a-d) Evolution of the magneto-resistance, $\Delta R/R(0)$, measured at 250 mK on Cr$_{1/3}$NbS$_2$ crystals of different thickness (see legends; blue/red curves represent data taken while sweeping $B$ in the direction indicated by the arrows). Resistance jumps and pronounced hysteresis are seen, absent in bulk Cr$_{1/3}$NbS$_2$ (see Fig. 1(d)). The insets zoom-in on a small $B$ range to make the smaller jumps visible. (e) Thickness dependence of the critical field $B_c$, corresponding to the highest value of $B$ at which a resistance jump is observed. The experimental data (dots) are in good agreement with calculations (solid line). The resistance continues to decrease for $B > B_c$, as shown in (f) by plotting $\frac{R(0.5 T)-R(B_c)}{R(0)}$ versus $t$. The data (dots) approximately exhibit a $1/(t-L_0)$ dependence (solid line) expected as the phenomenon originates from alignment of spins close to the crystal surfaces. The yellow regions in (e,f) denote the thickness range below $L_0$.}
\end{figure}

The relative change in resistance upon the application of an in-plane magnetic field, $\frac{\Delta R}{R(0)}\equiv\frac{R(B)-R(0)}{R(0)}$, is shown in Figs. 3(a-d) for several crystals containing  one ((a,b); $t=57$ nm and 79 nm respectively), two ((c); $t=113$ nm), and five ((d); $t=280$ nm) complete helix periods at $B=0$. The behavior is representative of what we observed in more than 10 devices realized with crystal having thickness between approximately 50 and 300 nm, exhibiting common features and systematic trends. For these crystals, hysteresis in the magneto-resistance upon reversing the sweeping direction of the applied field is always present, and is accompanied by sharp jumps. The phenomena cannot originate from magnetic domains, since all crystals are significantly thinner than the domain size~\cite{Togawa2015}. We find that the number of jumps tends to increase with increasing the crystal thickness, seemingly correlating with the number of complete periods present in the magnetic helix at $B=0$ (determined by $t/L_0$ where $L_0$ is the helix pitch). For instance, the crystals in Figs. 3(a) and 3(b) contain one full period and exhibit one jump, the crystal with $t=113$ nm (Fig. 3(c)) contains two full periods and exhibits two jumps. The $t=280$ nm crystal contains five full periods at $B=0$, and four jumps are observed, but the jump at $B\sim$ 0.17 T appears to be smeared, suggesting that it may originate from two jumps that are not resolved individually. The magnetic field at which the last jump occurs systematically increases with increasing crystal thickness, as shown in Fig. 3(e). Finally, the total change in the relative magneto-resistance measured after the last jump decreases with increasing thickness (see Fig. 3(f); this same quantity vanishes in bulk crystals).

All the observed trends can be understood in terms of a theoretical model known to properly describe the magnetic state of Cr$_{1/3}$NbS$_2$ (the model is discussed in several papers and here we only recall the key conceptual aspects; see also Supplemental Material~\cite{Support})~\cite{Togawa2012,Togawa2013,Chapman2014,Bornstein2015,Tsuruta2015,Borisov2009,Kishine2011}. The model enables the spin configuration to be determined through the minimization of the system energy expressed as a functional of the local magnetization. Its validity for Cr$_{1/3}$NbS$_2$ has been established by direct comparison with Lorentz microscopy experiments and magnetization measurements, which enable the model parameters to be extracted quantitatively~\cite{Togawa2012,Togawa2015}. The model has also been applied to strained MnSi thin-films~\cite{Wilson2013,Karhu2012}~\footnote{Bulk MnSi is not a chiral helical magnet, but MnSi thin-films are in a narrow range of thicknesses, due to strain from the substrate~\cite{Wilson2013,Karhu2012}.} to interpret magneto-transport data closely related to the ones discussed here. In this context, it was assumed that all changes causing a better aligned spin configuration result in smaller measured resistance~\cite{Support}.

In this same spirit, we reproduce some of the known results by solving the model numerically as a function of $B$ for crystals of different thicknesses, and we use the numerical solutions to illustrate the aspects of the behavior that account for the experimentally observed trends~\cite{Support}. Fig. 4(a-c) show the lowest energy spin configuration --represented by the $x$ component of the spin on the chromium atoms planes-- calculated at $B=0$ for crystals having $t=1.5$ $L_0$, 2.5 $L_0$, 5.5 $L_0$ (hosting respectively 1, 2, and 5 complete spin windings). For any given thickness, we calculate the spin configuration that minimizes the energy as the magnetic field is increased. This is done by minimizing the energy in each topological sector separately and comparing the resulting values to determine the absolute minimum. The outcome of this procedure is illustrated in Fig. 4(d) for the $t=2.5$ $L_0$ crystal. The lowest energy state at $B=0$ contains two magnetic solitons (Fig. 4(b)), but at a sufficiently large in-plane magnetic field the state with one soliton (Fig. 4(e)) becomes energetically favorable. At even larger $B$, the state with no solitons has the lowest energy, and all the spins are aligned, except those in proximity of the surfaces (Fig. 4(f)). Past this field, any further increase in $B$ only causes a gradual alignment of these spins. Such a sequence of transitions, with the number of solitons decreasing monotonously upon increasing $B$, is what the model predicts for all values of $t$ (not only for $t=2.5$ $L_0$).

\begin{figure}[t]
\includegraphics[width=8cm]{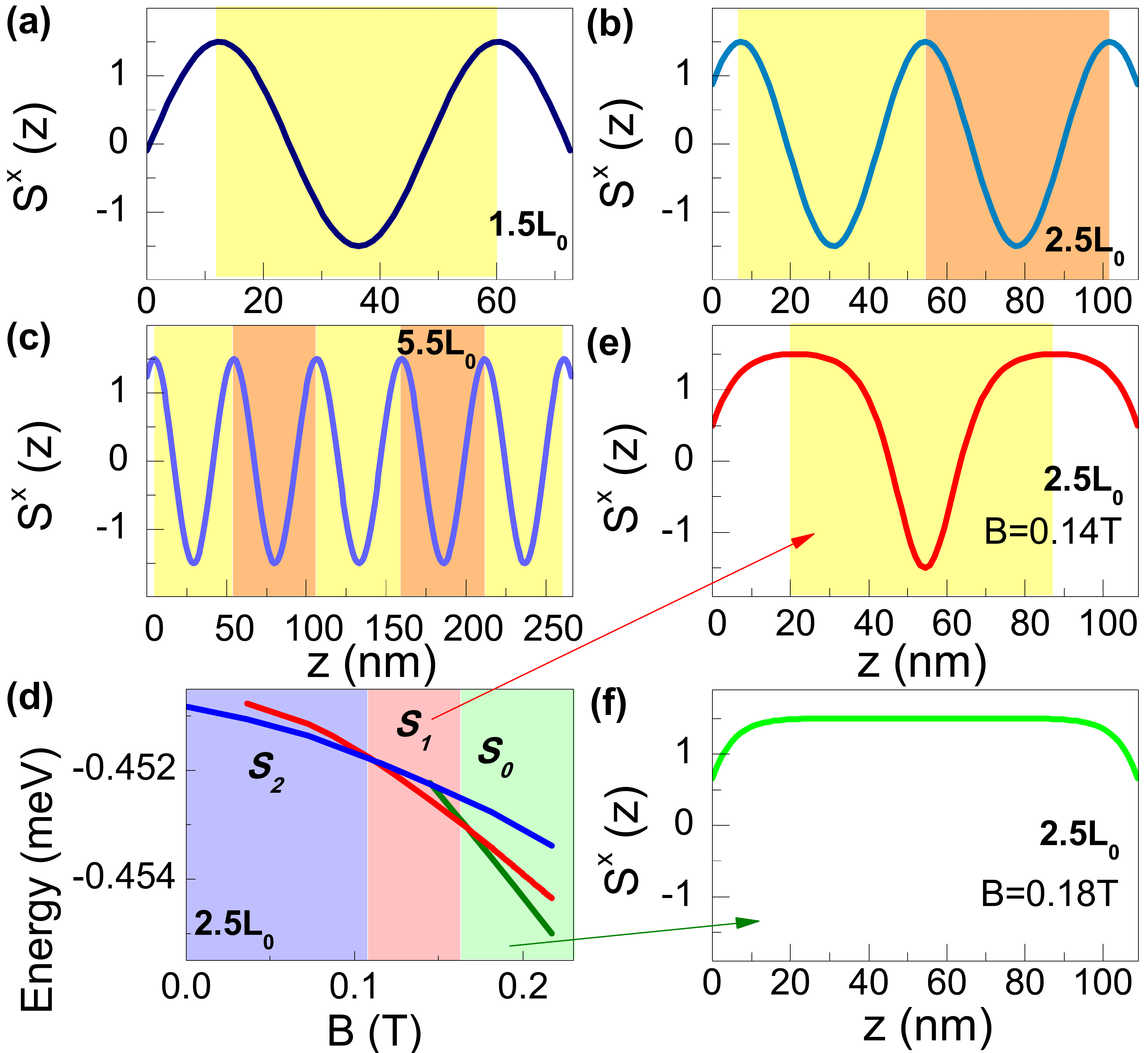}
\caption{\label{fig4} (Color online) (a-c) Lowest-energy configuration of the spin helix at $B=0$, represented by the $x$-component of spins on the Cr atoms ($S^x$), calculated using the model discussed in the main text for crystals with  $t=1.5$ $L_0$, 2.5 $L_0$, and 5.5 $L_0$ ($L_0=48$ nm). The colored areas mark complete spin windings, corresponding to the presence of one (a), two (b), and five (c) magnetic solitons. (d) The blue, red and green lines correspond to the lowest energy of configurations containing two, one and zero solitons, calculated as a function of $B$ for a crystal with $t=2.5$ $L_0$ . At low $B$ (blue shaded region) configurations containing two solitons (panel (c)) have the lowest energy; for intermediate values of $B$ (red shaded region) configurations with one soliton (panel (e)) have the lowest energy; for higher field configurations with no solitons (panel (f)) are energetically favorable.}
\end{figure}

This behavior accounts for all the experimentally observed trends. At each transition the resistance decreases, exhibiting a jump, because in configurations with a smaller number of solitons the spins are on average better aligned to the applied field~\cite{Support}. Therefore, the total number of jumps in the magneto-resistance equals the number of transition between states with different number of solitons, which can be calculated (and typically corresponds to the largest integer smaller than $t/L_0$). This is indeed in line with the experimental observations. Hysteresis is expected theoretically because configurations with number of solitons differing by one have different parity relative to the center of the crystals, so that whenever the system undergoes a transition, the spin configuration changes abruptly over the entire crystal. As a result, there is an energy barrier between solutions with different number of solitons, so that upon small variation of $B$ the system stays in the local minimum even when this is not the lowest energy configuration. The values of $B$ at which the transitions occur can be calculated as a function of $t$ and compared to the experiments. We focus on the last transition observed upon increasing $B$ past $B_c$ (\textit{i.e.}, the transition between the configurations containing 1 and 0 solitons), because this transition is present in all crystals with $t>50$ nm, giving us enough statistics for a meaningful quantitative comparison~\footnote{The $B$-values at which the transitions occur depend on thickness that --for exfoliated crystals-- cannot be deterministically controlled. This makes it difficult to compare the precise value of magnetic field for which a generic resistance jump is expected to occur with theory. Since devices of all thicknesses $t>L_0$ exhibit the jump corresponding to the transition to the ferromagnetic state, for this transition it is nevertheless possible to obtain enough data to make a statistically significant comparison.}. The data are systematic and the agreement with calculation is very good (see Fig. 3(e)). Finally, the model also explains the trend seen in the magnitude of the magneto-resistance measured upon varying $B$ past $B_c$. In this regime the magneto-resistance decreases because the spins next to the two crystals surfaces (see Fig. 4(f)) align progressively to the applied field as $B$ is increased. The gradual decrease in resistance is therefore a surface effect whose magnitude should be expected to decay approximately as $1/(t-L_0)$ and vanish in the bulk. This indeed agrees with the experiments (see Fig. 3(f), and Fig. 1(d) for the behavior of the bulk).

\begin{figure}[t]
\includegraphics[width=8.5cm]{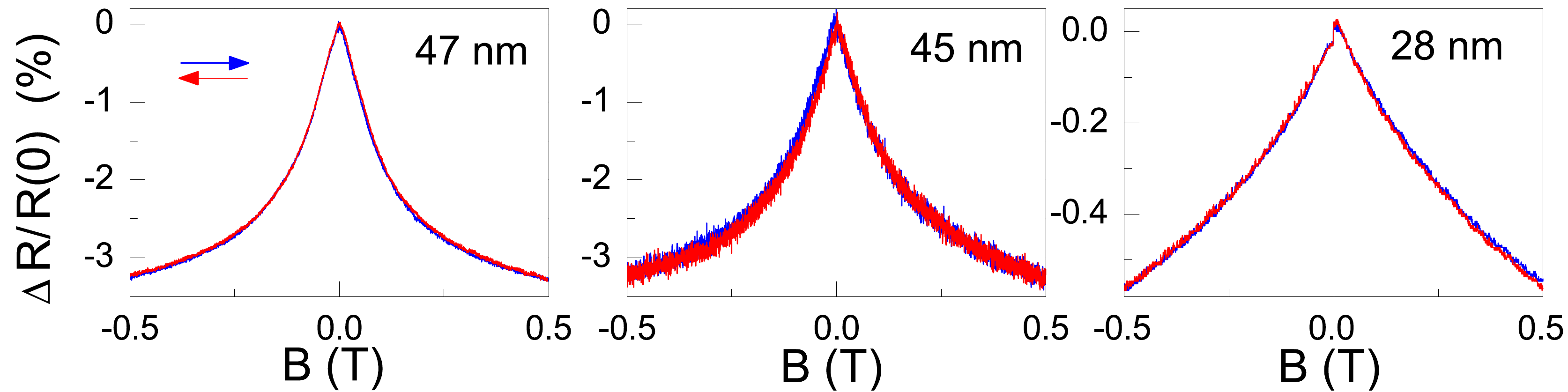}
\caption{\label{fig5} (Color online) Magneto-resistance $\Delta R/R(0)$ of crystals thinner than the helix pitch ($L_0=48$ nm) measured at 250 mK. The blue/red curves represent data measured while sweeping $B$ in opposite directions, as indicated by the arrows. In all cases, only a continuous decrease of the resistance is observed, and no jumps or hysteresis are present. }
\end{figure}

The relevance of these results stem from the fact that, contrary to all earlier work on Cr$_{1/3}$NbS$_2$~\cite{Togawa2015,Tsuruta2016a}, the crystals investigated here are substantially thinner than the magnetic domain size, so that domain switching can be excluded as possible origin of the phenomena. This and the agreement between observations and theoretically expected trends allow us to conclude that the observed jumps and the hysteresis do originate from transitions between states with different soliton number. The use of exfoliated crystals enables us to obtain further evidence supporting this conclusion, as it allows us to investigate crystals thinner than the helix pitch, a regime that has not been explored so far. No jumps and hysteresis in the magneto-resistance should be expected, because in this regime the soliton number vanishes for all values of $B$, and no topological transition can occur. Fig. 5 shows that this is indeed the case: in all Cr$_{1/3}$NbS$_2$ crystals thinner than the helix pitch at $B=0$ only a negative magneto-resistance is observed due to the gradual spin alignment upon increasing $B$. No jumps or hysteresis are present. This drastic qualitative difference in behavior observed upon changing the crystal thickness by only a few nanometers is striking. It provides conclusive evidence that in Cr$_{1/3}$NbS$_2$ the topological sector of the system can be controlled and probed by transport measurements.

In summary, we have performed experiments on very thin exfoliated crystals showing that the number of solitons in Cr$_{1/3}$NbS$_2$ can be controlled by selecting the appropriate crystal thickness and by acting on the applied field. The experiments further show that changes in the number of solitons manifest themselves in hysteretic magneto-resistance jumps. These results imply that the topological sector of the system can be deterministically controlled and probed by transport measurements.

L.W., D.-K.K., and A.F.M. are indebted to A. Ferreira for continuous technical assistance. N.C., F.M. and A.F.M. acknowledge financial support from the Swiss National Science Foundation. A.F.M. gratefully acknowledges financial support from the EU Graphene Flagship project. L.L. and D.M. acknowledge support from the National Science Foundation under grant No. DMR-1410428. F.L. and W.Z. are supported by LANL through LDRD program. I.M. acknowledges support from Department of Energy, Office of Basic Energy Science, Materials Science and Engineering Division. O.S.O supported by the Laboratory Directed Research and Development Program of Oak Ridge National Laboratory, managed by UT-Battelle, LLC, for the U. S. Department of Energy.


\begin{thebibliography}{31}%
\makeatletter
\providecommand \@ifxundefined [1]{%
 \@ifx{#1\undefined}
}%
\providecommand \@ifnum [1]{%
 \ifnum #1\expandafter \@firstoftwo
 \else \expandafter \@secondoftwo
 \fi
}%
\providecommand \@ifx [1]{%
 \ifx #1\expandafter \@firstoftwo
 \else \expandafter \@secondoftwo
 \fi
}%
\providecommand \natexlab [1]{#1}%
\providecommand \enquote  [1]{``#1''}%
\providecommand \bibnamefont  [1]{#1}%
\providecommand \bibfnamefont [1]{#1}%
\providecommand \citenamefont [1]{#1}%
\providecommand \href@noop [0]{\@secondoftwo}%
\providecommand \href [0]{\begingroup \@sanitize@url \@href}%
\providecommand \@href[1]{\@@startlink{#1}\@@href}%
\providecommand \@@href[1]{\endgroup#1\@@endlink}%
\providecommand \@sanitize@url [0]{\catcode `\\12\catcode `\$12\catcode
  `\&12\catcode `\#12\catcode `\^12\catcode `\_12\catcode `\%12\relax}%
\providecommand \@@startlink[1]{}%
\providecommand \@@endlink[0]{}%
\providecommand \url  [0]{\begingroup\@sanitize@url \@url }%
\providecommand \@url [1]{\endgroup\@href {#1}{\urlprefix }}%
\providecommand \urlprefix  [0]{URL }%
\providecommand \Eprint [0]{\href }%
\providecommand \doibase [0]{http://dx.doi.org/}%
\providecommand \selectlanguage [0]{\@gobble}%
\providecommand \bibinfo  [0]{\@secondoftwo}%
\providecommand \bibfield  [0]{\@secondoftwo}%
\providecommand \translation [1]{[#1]}%
\providecommand \BibitemOpen [0]{}%
\providecommand \bibitemStop [0]{}%
\providecommand \bibitemNoStop [0]{.\EOS\space}%
\providecommand \EOS [0]{\spacefactor3000\relax}%
\providecommand \BibitemShut  [1]{\csname bibitem#1\endcsname}%
\let\auto@bib@innerbib\@empty
\bibitem [{\citenamefont {Monastyrski\u{i}}(1993)}]{Monastyrski1993}%
  \BibitemOpen
  \bibfield  {author} {\bibinfo {author} {\bibfnamefont {M.~I.}\ \bibnamefont
  {Monastyrski\u{i}}},\ }\href@noop {} {\emph {\bibinfo {title} {Topology of
  gauge fields and condensed matter}}}\ (\bibinfo  {publisher} {Plenum Press},\
  \bibinfo {address} {New York},\ \bibinfo {year} {1993})\BibitemShut {NoStop}%
\bibitem [{\citenamefont {Volovik}(2003)}]{Volovik2003}%
  \BibitemOpen
  \bibfield  {author} {\bibinfo {author} {\bibfnamefont {G.~E.}\ \bibnamefont
  {Volovik}},\ }\href@noop {} {\emph {\bibinfo {title} {The universe in a
  helium droplet}}},\ International series of monographs on physics\ (\bibinfo
  {publisher} {Clarendon Press ; Oxford University Press},\ \bibinfo {address}
  {Oxford New York},\ \bibinfo {year} {2003})\BibitemShut {NoStop}%
\bibitem [{\citenamefont {Altland}\ and\ \citenamefont
  {Simons}(2010)}]{Altland2010}%
  \BibitemOpen
  \bibfield  {author} {\bibinfo {author} {\bibfnamefont {A.}~\bibnamefont
  {Altland}}\ and\ \bibinfo {author} {\bibfnamefont {B.}~\bibnamefont
  {Simons}},\ }\href@noop {} {\emph {\bibinfo {title} {Condensed matter field
  theory}}},\ \bibinfo {edition} {2nd}\ ed.\ (\bibinfo  {publisher} {Cambridge
  University Press},\ \bibinfo {address} {Cambridge ; New York},\ \bibinfo
  {year} {2010})\BibitemShut {NoStop}%
\bibitem [{\citenamefont {Fradkin}(2013)}]{Fradkin2013}%
  \BibitemOpen
  \bibfield  {author} {\bibinfo {author} {\bibfnamefont {E.}~\bibnamefont
  {Fradkin}},\ }\href@noop {} {\emph {\bibinfo {title} {Field theories of
  condensed matter physics}}},\ \bibinfo {edition} {2nd}\ ed.\ (\bibinfo
  {publisher} {Cambridge University Press},\ \bibinfo {address} {Cambridge},\
  \bibinfo {year} {2013})\BibitemShut {NoStop}%
\bibitem [{\citenamefont {Qi}\ and\ \citenamefont {Zhang}(2011)}]{Qi2011}%
  \BibitemOpen
  \bibfield  {author} {\bibinfo {author} {\bibfnamefont {X.-L.}\ \bibnamefont
  {Qi}}\ and\ \bibinfo {author} {\bibfnamefont {S.-C.}\ \bibnamefont {Zhang}},\
  }\href@noop {} {\bibfield  {journal} {\bibinfo  {journal} {Rev. Mod. Phys.}\
  }\textbf {\bibinfo {volume} {83}},\ \bibinfo {pages} {1057} (\bibinfo {year}
  {2011})}\BibitemShut {NoStop}%
\bibitem [{\citenamefont {Parkin}\ and\ \citenamefont
  {Friend}(1980{\natexlab{a}})}]{Parkin1980}%
  \BibitemOpen
  \bibfield  {author} {\bibinfo {author} {\bibfnamefont {S.~S.~P.}\
  \bibnamefont {Parkin}}\ and\ \bibinfo {author} {\bibfnamefont {R.~H.}\
  \bibnamefont {Friend}},\ }\href@noop {} {\bibfield  {journal} {\bibinfo
  {journal} {Phil. Mag. B}\ }\textbf {\bibinfo {volume} {41}},\ \bibinfo
  {pages} {65} (\bibinfo {year} {1980}{\natexlab{a}})}\BibitemShut {NoStop}%
\bibitem [{\citenamefont {Parkin}\ and\ \citenamefont
  {Friend}(1980{\natexlab{b}})}]{Parkin1980a}%
  \BibitemOpen
  \bibfield  {author} {\bibinfo {author} {\bibfnamefont {S.~S.~P.}\
  \bibnamefont {Parkin}}\ and\ \bibinfo {author} {\bibfnamefont {R.~H.}\
  \bibnamefont {Friend}},\ }\href@noop {} {\bibfield  {journal} {\bibinfo
  {journal} {Physica B+C}\ }\textbf {\bibinfo {volume} {99}},\ \bibinfo {pages}
  {219} (\bibinfo {year} {1980}{\natexlab{b}})}\BibitemShut {NoStop}%
\bibitem [{\citenamefont {Moriya}\ and\ \citenamefont
  {Miyadai}(1982)}]{Moriya1982}%
  \BibitemOpen
  \bibfield  {author} {\bibinfo {author} {\bibfnamefont {T.}~\bibnamefont
  {Moriya}}\ and\ \bibinfo {author} {\bibfnamefont {T.}~\bibnamefont
  {Miyadai}},\ }\href@noop {} {\bibfield  {journal} {\bibinfo  {journal} {Solid
  State Commun.}\ }\textbf {\bibinfo {volume} {42}},\ \bibinfo {pages} {209}
  (\bibinfo {year} {1982})}\BibitemShut {NoStop}%
\bibitem [{\citenamefont {Miyadai}\ \emph {et~al.}(1983)\citenamefont
  {Miyadai}, \citenamefont {Kikuchi}, \citenamefont {Kondo}, \citenamefont
  {Sakka}, \citenamefont {Arai},\ and\ \citenamefont {Ishikawa}}]{Miyadai1983}%
  \BibitemOpen
  \bibfield  {author} {\bibinfo {author} {\bibfnamefont {T.}~\bibnamefont
  {Miyadai}}, \bibinfo {author} {\bibfnamefont {K.}~\bibnamefont {Kikuchi}},
  \bibinfo {author} {\bibfnamefont {H.}~\bibnamefont {Kondo}}, \bibinfo
  {author} {\bibfnamefont {S.}~\bibnamefont {Sakka}}, \bibinfo {author}
  {\bibfnamefont {M.}~\bibnamefont {Arai}}, \ and\ \bibinfo {author}
  {\bibfnamefont {Y.}~\bibnamefont {Ishikawa}},\ }\href@noop {} {\bibfield
  {journal} {\bibinfo  {journal} {J. Phys. Soc. Jpn.}\ }\textbf {\bibinfo
  {volume} {52}},\ \bibinfo {pages} {1394} (\bibinfo {year}
  {1983})}\BibitemShut {NoStop}%
\bibitem [{\citenamefont {Yurii}(1984)}]{Yurii1984}%
  \BibitemOpen
  \bibfield  {author} {\bibinfo {author} {\bibfnamefont {A.~I.}\ \bibnamefont
  {Yurii}},\ }\href@noop {} {\bibfield  {journal} {\bibinfo  {journal} {Sov.
  Phys. Uspekhi}\ }\textbf {\bibinfo {volume} {27}},\ \bibinfo {pages} {845}
  (\bibinfo {year} {1984})}\BibitemShut {NoStop}%
\bibitem [{\citenamefont {Kousaka}\ \emph {et~al.}(2009)\citenamefont
  {Kousaka}, \citenamefont {Nakao}, \citenamefont {Kishine}, \citenamefont
  {Akita}, \citenamefont {Inoue},\ and\ \citenamefont
  {Akimitsu}}]{Kousaka2009}%
  \BibitemOpen
  \bibfield  {author} {\bibinfo {author} {\bibfnamefont {Y.}~\bibnamefont
  {Kousaka}}, \bibinfo {author} {\bibfnamefont {Y.}~\bibnamefont {Nakao}},
  \bibinfo {author} {\bibfnamefont {J.}~\bibnamefont {Kishine}}, \bibinfo
  {author} {\bibfnamefont {M.}~\bibnamefont {Akita}}, \bibinfo {author}
  {\bibfnamefont {K.}~\bibnamefont {Inoue}}, \ and\ \bibinfo {author}
  {\bibfnamefont {J.}~\bibnamefont {Akimitsu}},\ }\href@noop {} {\bibfield
  {journal} {\bibinfo  {journal} {Nuc. Ins. Met. Phys. Res. Sec. A}\ }\textbf
  {\bibinfo {volume} {600}},\ \bibinfo {pages} {250} (\bibinfo {year}
  {2009})}\BibitemShut {NoStop}%
\bibitem [{\citenamefont {Togawa}\ \emph {et~al.}(2012)\citenamefont {Togawa},
  \citenamefont {Koyama}, \citenamefont {Takayanagi}, \citenamefont {Mori},
  \citenamefont {Kousaka}, \citenamefont {Akimitsu}, \citenamefont {Nishihara},
  \citenamefont {Inoue}, \citenamefont {Ovchinnikov},\ and\ \citenamefont
  {Kishine}}]{Togawa2012}%
  \BibitemOpen
  \bibfield  {author} {\bibinfo {author} {\bibfnamefont {Y.}~\bibnamefont
  {Togawa}}, \bibinfo {author} {\bibfnamefont {T.}~\bibnamefont {Koyama}},
  \bibinfo {author} {\bibfnamefont {K.}~\bibnamefont {Takayanagi}}, \bibinfo
  {author} {\bibfnamefont {S.}~\bibnamefont {Mori}}, \bibinfo {author}
  {\bibfnamefont {Y.}~\bibnamefont {Kousaka}}, \bibinfo {author} {\bibfnamefont
  {J.}~\bibnamefont {Akimitsu}}, \bibinfo {author} {\bibfnamefont
  {S.}~\bibnamefont {Nishihara}}, \bibinfo {author} {\bibfnamefont
  {K.}~\bibnamefont {Inoue}}, \bibinfo {author} {\bibfnamefont {A.~S.}\
  \bibnamefont {Ovchinnikov}}, \ and\ \bibinfo {author} {\bibfnamefont
  {J.}~\bibnamefont {Kishine}},\ }\href@noop {} {\bibfield  {journal} {\bibinfo
   {journal} {Phys. Rev. Lett.}\ }\textbf {\bibinfo {volume} {108}},\ \bibinfo
  {pages} {107202} (\bibinfo {year} {2012})}\BibitemShut {NoStop}%
\bibitem [{\citenamefont {Ghimire}\ \emph {et~al.}(2013)\citenamefont
  {Ghimire}, \citenamefont {McGuire}, \citenamefont {Parker}, \citenamefont
  {Sipos}, \citenamefont {Tang}, \citenamefont {Yan}, \citenamefont {Sales},\
  and\ \citenamefont {Mandrus}}]{Ghimire2013}%
  \BibitemOpen
  \bibfield  {author} {\bibinfo {author} {\bibfnamefont {N.~J.}\ \bibnamefont
  {Ghimire}}, \bibinfo {author} {\bibfnamefont {M.~A.}\ \bibnamefont
  {McGuire}}, \bibinfo {author} {\bibfnamefont {D.~S.}\ \bibnamefont {Parker}},
  \bibinfo {author} {\bibfnamefont {B.}~\bibnamefont {Sipos}}, \bibinfo
  {author} {\bibfnamefont {S.}~\bibnamefont {Tang}}, \bibinfo {author}
  {\bibfnamefont {J.~Q.}\ \bibnamefont {Yan}}, \bibinfo {author} {\bibfnamefont
  {B.~C.}\ \bibnamefont {Sales}}, \ and\ \bibinfo {author} {\bibfnamefont
  {D.}~\bibnamefont {Mandrus}},\ }\href@noop {} {\bibfield  {journal} {\bibinfo
   {journal} {Phys. Rev. B}\ }\textbf {\bibinfo {volume} {87}},\ \bibinfo
  {pages} {104403} (\bibinfo {year} {2013})}\BibitemShut {NoStop}%
\bibitem [{\citenamefont {Togawa}\ \emph {et~al.}(2013)\citenamefont {Togawa},
  \citenamefont {Kousaka}, \citenamefont {Nishihara}, \citenamefont {Inoue},
  \citenamefont {Akimitsu}, \citenamefont {Ovchinnikov},\ and\ \citenamefont
  {Kishine}}]{Togawa2013}%
  \BibitemOpen
  \bibfield  {author} {\bibinfo {author} {\bibfnamefont {Y.}~\bibnamefont
  {Togawa}}, \bibinfo {author} {\bibfnamefont {Y.}~\bibnamefont {Kousaka}},
  \bibinfo {author} {\bibfnamefont {S.}~\bibnamefont {Nishihara}}, \bibinfo
  {author} {\bibfnamefont {K.}~\bibnamefont {Inoue}}, \bibinfo {author}
  {\bibfnamefont {J.}~\bibnamefont {Akimitsu}}, \bibinfo {author}
  {\bibfnamefont {A.~S.}\ \bibnamefont {Ovchinnikov}}, \ and\ \bibinfo {author}
  {\bibfnamefont {J.}~\bibnamefont {Kishine}},\ }\href@noop {} {\bibfield
  {journal} {\bibinfo  {journal} {Phys. Rev. Lett.}\ }\textbf {\bibinfo
  {volume} {111}},\ \bibinfo {pages} {197204} (\bibinfo {year}
  {2013})}\BibitemShut {NoStop}%
\bibitem [{\citenamefont {Chapman}\ \emph {et~al.}(2014)\citenamefont
  {Chapman}, \citenamefont {Bornstein}, \citenamefont {Ghimire}, \citenamefont
  {Mandrus},\ and\ \citenamefont {Lee}}]{Chapman2014}%
  \BibitemOpen
  \bibfield  {author} {\bibinfo {author} {\bibfnamefont {B.~J.}\ \bibnamefont
  {Chapman}}, \bibinfo {author} {\bibfnamefont {A.~C.}\ \bibnamefont
  {Bornstein}}, \bibinfo {author} {\bibfnamefont {N.~J.}\ \bibnamefont
  {Ghimire}}, \bibinfo {author} {\bibfnamefont {D.}~\bibnamefont {Mandrus}}, \
  and\ \bibinfo {author} {\bibfnamefont {M.}~\bibnamefont {Lee}},\ }\href@noop
  {} {\bibfield  {journal} {\bibinfo  {journal} {Appl. Phys. Lett.}\ }\textbf
  {\bibinfo {volume} {105}},\ \bibinfo {pages} {072405} (\bibinfo {year}
  {2014})}\BibitemShut {NoStop}%
\bibitem [{\citenamefont {Bornstein}\ \emph {et~al.}(2015)\citenamefont
  {Bornstein}, \citenamefont {Chapman}, \citenamefont {Ghimire}, \citenamefont
  {Mandrus}, \citenamefont {Parker},\ and\ \citenamefont
  {Lee}}]{Bornstein2015}%
  \BibitemOpen
  \bibfield  {author} {\bibinfo {author} {\bibfnamefont {A.~C.}\ \bibnamefont
  {Bornstein}}, \bibinfo {author} {\bibfnamefont {B.~J.}\ \bibnamefont
  {Chapman}}, \bibinfo {author} {\bibfnamefont {N.~J.}\ \bibnamefont
  {Ghimire}}, \bibinfo {author} {\bibfnamefont {D.~G.}\ \bibnamefont
  {Mandrus}}, \bibinfo {author} {\bibfnamefont {D.~S.}\ \bibnamefont {Parker}},
  \ and\ \bibinfo {author} {\bibfnamefont {M.}~\bibnamefont {Lee}},\
  }\href@noop {} {\bibfield  {journal} {\bibinfo  {journal} {Phys. Rev. B}\
  }\textbf {\bibinfo {volume} {91}},\ \bibinfo {pages} {184401} (\bibinfo
  {year} {2015})}\BibitemShut {NoStop}%
\bibitem [{\citenamefont {Tsuruta}\ \emph
  {et~al.}(2016{\natexlab{a}})\citenamefont {Tsuruta}, \citenamefont {Mito},
  \citenamefont {Deguchi}, \citenamefont {Kishine}, \citenamefont {Kousaka},
  \citenamefont {Akimitsu},\ and\ \citenamefont {Inoue}}]{Tsuruta2016}%
  \BibitemOpen
  \bibfield  {author} {\bibinfo {author} {\bibfnamefont {K.}~\bibnamefont
  {Tsuruta}}, \bibinfo {author} {\bibfnamefont {M.}~\bibnamefont {Mito}},
  \bibinfo {author} {\bibfnamefont {H.}~\bibnamefont {Deguchi}}, \bibinfo
  {author} {\bibfnamefont {J.}~\bibnamefont {Kishine}}, \bibinfo {author}
  {\bibfnamefont {Y.}~\bibnamefont {Kousaka}}, \bibinfo {author} {\bibfnamefont
  {J.}~\bibnamefont {Akimitsu}}, \ and\ \bibinfo {author} {\bibfnamefont
  {K.}~\bibnamefont {Inoue}},\ }\href@noop {} {\bibfield  {journal} {\bibinfo
  {journal} {Phys. Rev. B}\ }\textbf {\bibinfo {volume} {93}},\ \bibinfo
  {pages} {104402} (\bibinfo {year} {2016}{\natexlab{a}})}\BibitemShut
  {NoStop}%
\bibitem [{\citenamefont {Dzyaloshinskii}(1958)}]{Dzyaloshinskii1958}%
  \BibitemOpen
  \bibfield  {author} {\bibinfo {author} {\bibfnamefont {I.~E.}\ \bibnamefont
  {Dzyaloshinskii}},\ }\href@noop {} {\bibfield  {journal} {\bibinfo  {journal}
  {J. Phys. Chem. Solids}\ }\textbf {\bibinfo {volume} {4}},\ \bibinfo {pages}
  {241} (\bibinfo {year} {1958})}\BibitemShut {NoStop}%
\bibitem [{\citenamefont {Moriya}(1960)}]{Moriya1960}%
  \BibitemOpen
  \bibfield  {author} {\bibinfo {author} {\bibfnamefont {T.}~\bibnamefont
  {Moriya}},\ }\href@noop {} {\bibfield  {journal} {\bibinfo  {journal} {Phys.
  Rev.}\ }\textbf {\bibinfo {volume} {120}},\ \bibinfo {pages} {91} (\bibinfo
  {year} {1960})}\BibitemShut {NoStop}%
\bibitem [{\citenamefont {Kishine}\ \emph {et~al.}(2014)\citenamefont
  {Kishine}, \citenamefont {Bostrem}, \citenamefont {Ovchinnikov},\ and\
  \citenamefont {Sinitsyn}}]{Kishine2014}%
  \BibitemOpen
  \bibfield  {author} {\bibinfo {author} {\bibfnamefont {J.}~\bibnamefont
  {Kishine}}, \bibinfo {author} {\bibfnamefont {I.~G.}\ \bibnamefont
  {Bostrem}}, \bibinfo {author} {\bibfnamefont {A.~S.}\ \bibnamefont
  {Ovchinnikov}}, \ and\ \bibinfo {author} {\bibfnamefont {V.~E.}\ \bibnamefont
  {Sinitsyn}},\ }\href@noop {} {\bibfield  {journal} {\bibinfo  {journal}
  {Phys. Rev. B}\ }\textbf {\bibinfo {volume} {89}},\ \bibinfo {pages} {014419}
  (\bibinfo {year} {2014})}\BibitemShut {NoStop}%
\bibitem [{\citenamefont {Tsuruta}\ \emph {et~al.}(2015)\citenamefont
  {Tsuruta}, \citenamefont {Mito}, \citenamefont {Kousaka}, \citenamefont
  {Akimitsu}, \citenamefont {Kishine}, \citenamefont {Togawa}, \citenamefont
  {Ohsumi},\ and\ \citenamefont {Inoue}}]{Tsuruta2015}%
  \BibitemOpen
  \bibfield  {author} {\bibinfo {author} {\bibfnamefont {K.}~\bibnamefont
  {Tsuruta}}, \bibinfo {author} {\bibfnamefont {M.}~\bibnamefont {Mito}},
  \bibinfo {author} {\bibfnamefont {Y.}~\bibnamefont {Kousaka}}, \bibinfo
  {author} {\bibfnamefont {J.}~\bibnamefont {Akimitsu}}, \bibinfo {author}
  {\bibfnamefont {J.}~\bibnamefont {Kishine}}, \bibinfo {author} {\bibfnamefont
  {Y.}~\bibnamefont {Togawa}}, \bibinfo {author} {\bibfnamefont
  {H.}~\bibnamefont {Ohsumi}}, \ and\ \bibinfo {author} {\bibfnamefont
  {K.}~\bibnamefont {Inoue}},\ }\href@noop {} {\bibfield  {journal} {\bibinfo
  {journal} {J. Phys. Soc. Jpn.}\ }\textbf {\bibinfo {volume} {85}},\ \bibinfo
  {pages} {013707} (\bibinfo {year} {2015})}\BibitemShut {NoStop}%
\bibitem [{\citenamefont {Togawa}\ \emph {et~al.}(2015)\citenamefont {Togawa},
  \citenamefont {Koyama}, \citenamefont {Nishimori}, \citenamefont {Matsumoto},
  \citenamefont {McVitie}, \citenamefont {McGrouther}, \citenamefont {Stamps},
  \citenamefont {Kousaka}, \citenamefont {Akimitsu}, \citenamefont {Nishihara},
  \citenamefont {Inoue}, \citenamefont {Bostrem}, \citenamefont {Sinitsyn},
  \citenamefont {Ovchinnikov},\ and\ \citenamefont {Kishine}}]{Togawa2015}%
  \BibitemOpen
  \bibfield  {author} {\bibinfo {author} {\bibfnamefont {Y.}~\bibnamefont
  {Togawa}}, \bibinfo {author} {\bibfnamefont {T.}~\bibnamefont {Koyama}},
  \bibinfo {author} {\bibfnamefont {Y.}~\bibnamefont {Nishimori}}, \bibinfo
  {author} {\bibfnamefont {Y.}~\bibnamefont {Matsumoto}}, \bibinfo {author}
  {\bibfnamefont {S.}~\bibnamefont {McVitie}}, \bibinfo {author} {\bibfnamefont
  {D.}~\bibnamefont {McGrouther}}, \bibinfo {author} {\bibfnamefont {R.~L.}\
  \bibnamefont {Stamps}}, \bibinfo {author} {\bibfnamefont {Y.}~\bibnamefont
  {Kousaka}}, \bibinfo {author} {\bibfnamefont {J.}~\bibnamefont {Akimitsu}},
  \bibinfo {author} {\bibfnamefont {S.}~\bibnamefont {Nishihara}}, \bibinfo
  {author} {\bibfnamefont {K.}~\bibnamefont {Inoue}}, \bibinfo {author}
  {\bibfnamefont {I.~G.}\ \bibnamefont {Bostrem}}, \bibinfo {author}
  {\bibfnamefont {V.~E.}\ \bibnamefont {Sinitsyn}}, \bibinfo {author}
  {\bibfnamefont {A.~S.}\ \bibnamefont {Ovchinnikov}}, \ and\ \bibinfo {author}
  {\bibfnamefont {J.}~\bibnamefont {Kishine}},\ }\href@noop {} {\bibfield
  {journal} {\bibinfo  {journal} {Phys. Rev. B}\ }\textbf {\bibinfo {volume}
  {92}},\ \bibinfo {pages} {220412} (\bibinfo {year} {2015})}\BibitemShut
  {NoStop}%
\bibitem [{\citenamefont {Tsuruta}\ \emph
  {et~al.}(2016{\natexlab{b}})\citenamefont {Tsuruta}, \citenamefont {Mito},
  \citenamefont {Kousaka}, \citenamefont {Akimitsu}, \citenamefont {Kishine},
  \citenamefont {Togawa},\ and\ \citenamefont {Inoue}}]{Tsuruta2016a}%
  \BibitemOpen
  \bibfield  {author} {\bibinfo {author} {\bibfnamefont {K.}~\bibnamefont
  {Tsuruta}}, \bibinfo {author} {\bibfnamefont {M.}~\bibnamefont {Mito}},
  \bibinfo {author} {\bibfnamefont {Y.}~\bibnamefont {Kousaka}}, \bibinfo
  {author} {\bibfnamefont {J.}~\bibnamefont {Akimitsu}}, \bibinfo {author}
  {\bibfnamefont {J.}~\bibnamefont {Kishine}}, \bibinfo {author} {\bibfnamefont
  {Y.}~\bibnamefont {Togawa}}, \ and\ \bibinfo {author} {\bibfnamefont
  {K.}~\bibnamefont {Inoue}},\ }\href@noop {} {\bibfield  {journal} {\bibinfo
  {journal} {J. Appl. Phys.}\ }\textbf {\bibinfo {volume} {120}},\ \bibinfo
  {pages} {5} (\bibinfo {year} {2016}{\natexlab{b}})}\BibitemShut {NoStop}%
\bibitem [{\citenamefont {Bak}\ and\ \citenamefont {Jensen}(1980)}]{Bak1980}%
  \BibitemOpen
  \bibfield  {author} {\bibinfo {author} {\bibfnamefont {P.}~\bibnamefont
  {Bak}}\ and\ \bibinfo {author} {\bibfnamefont {M.~H.}\ \bibnamefont
  {Jensen}},\ }\href@noop {} {\bibfield  {journal} {\bibinfo  {journal} {J.
  Phys. C}\ }\textbf {\bibinfo {volume} {13}},\ \bibinfo {pages} {L881}
  (\bibinfo {year} {1980})}\BibitemShut {NoStop}%
\bibitem [{Sup()}]{Support}%
  \BibitemOpen
  \href@noop {} {\ }\bibinfo {note} {See Supplemental Material at [URL will be
  inserted by publisher] for a detailed description of our model, the values of
  parameters used for the numerical calculations, and the results.}\BibitemShut
  {Stop}%
\bibitem [{\citenamefont {Borisov}\ \emph {et~al.}(2009)\citenamefont
  {Borisov}, \citenamefont {Kishine}, \citenamefont {Bostrem},\ and\
  \citenamefont {Ovchinnikov}}]{Borisov2009}%
  \BibitemOpen
  \bibfield  {author} {\bibinfo {author} {\bibfnamefont {A.~B.}\ \bibnamefont
  {Borisov}}, \bibinfo {author} {\bibfnamefont {J.}~\bibnamefont {Kishine}},
  \bibinfo {author} {\bibfnamefont {I.~G.}\ \bibnamefont {Bostrem}}, \ and\
  \bibinfo {author} {\bibfnamefont {A.~S.}\ \bibnamefont {Ovchinnikov}},\
  }\href@noop {} {\bibfield  {journal} {\bibinfo  {journal} {Phys. Rev. B}\
  }\textbf {\bibinfo {volume} {79}},\ \bibinfo {pages} {134436} (\bibinfo
  {year} {2009})}\BibitemShut {NoStop}%
\bibitem [{\citenamefont {Kishine}\ \emph {et~al.}(2011)\citenamefont
  {Kishine}, \citenamefont {Proskurin},\ and\ \citenamefont
  {Ovchinnikov}}]{Kishine2011}%
  \BibitemOpen
  \bibfield  {author} {\bibinfo {author} {\bibfnamefont {J.}~\bibnamefont
  {Kishine}}, \bibinfo {author} {\bibfnamefont {I.~V.}\ \bibnamefont
  {Proskurin}}, \ and\ \bibinfo {author} {\bibfnamefont {A.~S.}\ \bibnamefont
  {Ovchinnikov}},\ }\href@noop {} {\bibfield  {journal} {\bibinfo  {journal}
  {Phys. Rev. Lett.}\ }\textbf {\bibinfo {volume} {107}},\ \bibinfo {pages}
  {017205} (\bibinfo {year} {2011})}\BibitemShut {NoStop}%
\bibitem [{\citenamefont {Wilson}\ \emph {et~al.}(2013)\citenamefont {Wilson},
  \citenamefont {Karhu}, \citenamefont {Lake}, \citenamefont {Quigley},
  \citenamefont {Meynell}, \citenamefont {Bogdanov}, \citenamefont {Fritzsche},
  \citenamefont {Rö{\ss}ler},\ and\ \citenamefont {Monchesky}}]{Wilson2013}%
  \BibitemOpen
  \bibfield  {author} {\bibinfo {author} {\bibfnamefont {M.~N.}\ \bibnamefont
  {Wilson}}, \bibinfo {author} {\bibfnamefont {E.~A.}\ \bibnamefont {Karhu}},
  \bibinfo {author} {\bibfnamefont {D.~P.}\ \bibnamefont {Lake}}, \bibinfo
  {author} {\bibfnamefont {A.~S.}\ \bibnamefont {Quigley}}, \bibinfo {author}
  {\bibfnamefont {S.}~\bibnamefont {Meynell}}, \bibinfo {author} {\bibfnamefont
  {A.~N.}\ \bibnamefont {Bogdanov}}, \bibinfo {author} {\bibfnamefont
  {H.}~\bibnamefont {Fritzsche}}, \bibinfo {author} {\bibfnamefont {U.~K.}\
  \bibnamefont {Rö{\ss}ler}}, \ and\ \bibinfo {author} {\bibfnamefont {T.~L.}\
  \bibnamefont {Monchesky}},\ }\href@noop {} {\bibfield  {journal} {\bibinfo
  {journal} {Phys. Rev. B}\ }\textbf {\bibinfo {volume} {88}},\ \bibinfo
  {pages} {214420} (\bibinfo {year} {2013})}\BibitemShut {NoStop}%
\bibitem [{\citenamefont {Karhu}\ \emph {et~al.}(2012)\citenamefont {Karhu},
  \citenamefont {Rö{\ss}ler}, \citenamefont {Bogdanov}, \citenamefont
  {Kahwaji}, \citenamefont {Kirby}, \citenamefont {Fritzsche}, \citenamefont
  {Robertson}, \citenamefont {Majkrzak},\ and\ \citenamefont
  {Monchesky}}]{Karhu2012}%
  \BibitemOpen
  \bibfield  {author} {\bibinfo {author} {\bibfnamefont {E.~A.}\ \bibnamefont
  {Karhu}}, \bibinfo {author} {\bibfnamefont {U.~K.}\ \bibnamefont
  {Rö{\ss}ler}}, \bibinfo {author} {\bibfnamefont {A.~N.}\ \bibnamefont
  {Bogdanov}}, \bibinfo {author} {\bibfnamefont {S.}~\bibnamefont {Kahwaji}},
  \bibinfo {author} {\bibfnamefont {B.~J.}\ \bibnamefont {Kirby}}, \bibinfo
  {author} {\bibfnamefont {H.}~\bibnamefont {Fritzsche}}, \bibinfo {author}
  {\bibfnamefont {M.~D.}\ \bibnamefont {Robertson}}, \bibinfo {author}
  {\bibfnamefont {C.~F.}\ \bibnamefont {Majkrzak}}, \ and\ \bibinfo {author}
  {\bibfnamefont {T.~L.}\ \bibnamefont {Monchesky}},\ }\href@noop {} {\bibfield
   {journal} {\bibinfo  {journal} {Phys. Rev. B}\ }\textbf {\bibinfo {volume}
  {85}},\ \bibinfo {pages} {094429} (\bibinfo {year} {2012})}\BibitemShut
  {NoStop}%
\bibitem [{Note1()}]{Note1}%
  \BibitemOpen
  \bibinfo {note} {Bulk MnSi is not a chiral helical magnet, but MnSi
  thin-films are in a narrow range of thicknesses, due to strain from the
  substrate~\cite {Wilson2013,Karhu2012}.}\BibitemShut {Stop}%
\bibitem [{Note2()}]{Note2}%
  \BibitemOpen
  \bibinfo {note} {The $B$-values at which the transitions occur depend on
  thickness that --for exfoliated crystals-- cannot be deterministically
  controlled. This makes it difficult to compare the precise value of magnetic
  field for which a generic resistance jump is expected to occur with theory.
  Since devices of all thicknesses $t>L_0$ exhibit the jump corresponding to
  the transition to the ferromagnetic state, for this transition it is
  nevertheless possible to obtain enough data to make a statistically
  significant comparison.}\BibitemShut {Stop}%
\end{thebibliography}
%

\begin{center}
\textbf{\large Supplemental Materials}
\end{center}

\newcommand{\bhy}{\hat y}
\newcommand{\bhz}{ \hat z}
\newcommand{\bp}{\mathbf p}
\newcommand{\bb}{\mathbf b}
\newcommand{\bc}{\mathbf c}
\newcommand{\bkp}{{\mathbf k}_{\parallel}}
\newcommand{\nn}{\nonumber}
\newcommand{\ft}{\footnote{\tiny}}
\newcommand{\sgn}{\operatorname{sgn}}
\newcommand{\bk}{\mathbf{k}}
\newcommand{\disav}[1]{\left\langle #1\right\rangle}
\newcommand{\bu}{\mathbf{u}}
\newcommand{\bs}{\mathbf{s}}
\newcommand{\ba}{\mathbf{a}}
\newcommand{\bmm}{\mathbf{m}}
\newcommand{\bM}{\mathbf{M}}
\newcommand{\bP}{\mathbf{P}}
\newcommand{\bD}{\mathbf{D}}
\newcommand{\bR}{\mathbf{R}}
\newcommand{\bG}{\mathbf{G}}
\newcommand{\cR}{{\cal R}}
\newcommand{\cS}{{\cal S}}
\newcommand{\cV}{{\cal V}}
\newcommand{\bK} {{\bf{K}}}
\newcommand{\bA} {{\bf{A}}}
\newcommand{\cD} {{\cal D}}
\newcommand{\bhn}{\bf{\hat n}}
\newcommand{\bhr}{\mathbf{\hat r}}
\newcommand{\bnabla}{\mathbf\nabla}
\newcommand{\br}{\mathbf r}
\newcommand{\bh}{\mathbf h}
\newcommand{\bq}{\mathbf q}
\newcommand{\bx}{\mathbf r}
\newcommand{\by}{\mathbf y}
\newcommand{\bz}{\mathbf z}
\newcommand{\bn}{\mathbf n}
\newcommand{\bj}{\mathbf j}
\newcommand{\bpi}{\bm \pi}
\newcommand{\hn}{\hat n}
\newcommand{\dg}{\dagger}
\newcommand{\bS}{\mathbf{S}}
\newcommand{\bmu} {\bm \mu}
\newcommand{\bchi} {\bm \chi}
\newcommand{\brho} {\bm \rho}

\newcommand{\be}{\begin{eqnarray}}
\newcommand{\ee}{\end{eqnarray}}
\newcommand{\la}{\langle}
\newcommand{\ra}{\rangle}
\newcommand{\rar}{\rightarrow}
\newcommand{\da}{\downarrow}
\newcommand{\ua}{\uparrow}
\newcommand{\comment}[1]{\textcolor{red}{$\rhd$ \texttt{#1} $\lhd$}}

\section{Numerical results}

The Cr$_{1/3}$NbS$_2$ is effectively described by the following microscopic Hamiltonian:

\begin{equation}
H=\sum_{i}-J{\bf S}_i\cdot{\bf S}_{i+1}-{\bf D}\cdot ({\bf S}_i\times{\bf S}_{i+1})+A_z(S^z_i)^2-\mu_B{\bf B} \cdot {\bf S}_i, \label{eq:H}
\end{equation}
where ${\bf D}=(0,0,-0.016J)$ is the coupling constant of the Dzyaloshinsky-Moriya interaction, $A_z=0.10J$ corresponds to the magnetocrystalline  anisotropy and ${\bf B}$ is the external magnetic field. The ratios $D/J$ and $A_z/J$ are taken from Ref. [15]. The ferromagnetic coupling constant $J$ has been adjusted to $J=0.209\mathrm{meV}$, so that the numerical value of the critical field $\mu_B B^c/J=0.02356$ in the thermodynamic limit matches the critical field for the bulk crystal $B^c\approx0.17T$ (see Fig.1(d) of the main text).
\begin{figure}[h]
\includegraphics[width=8cm]{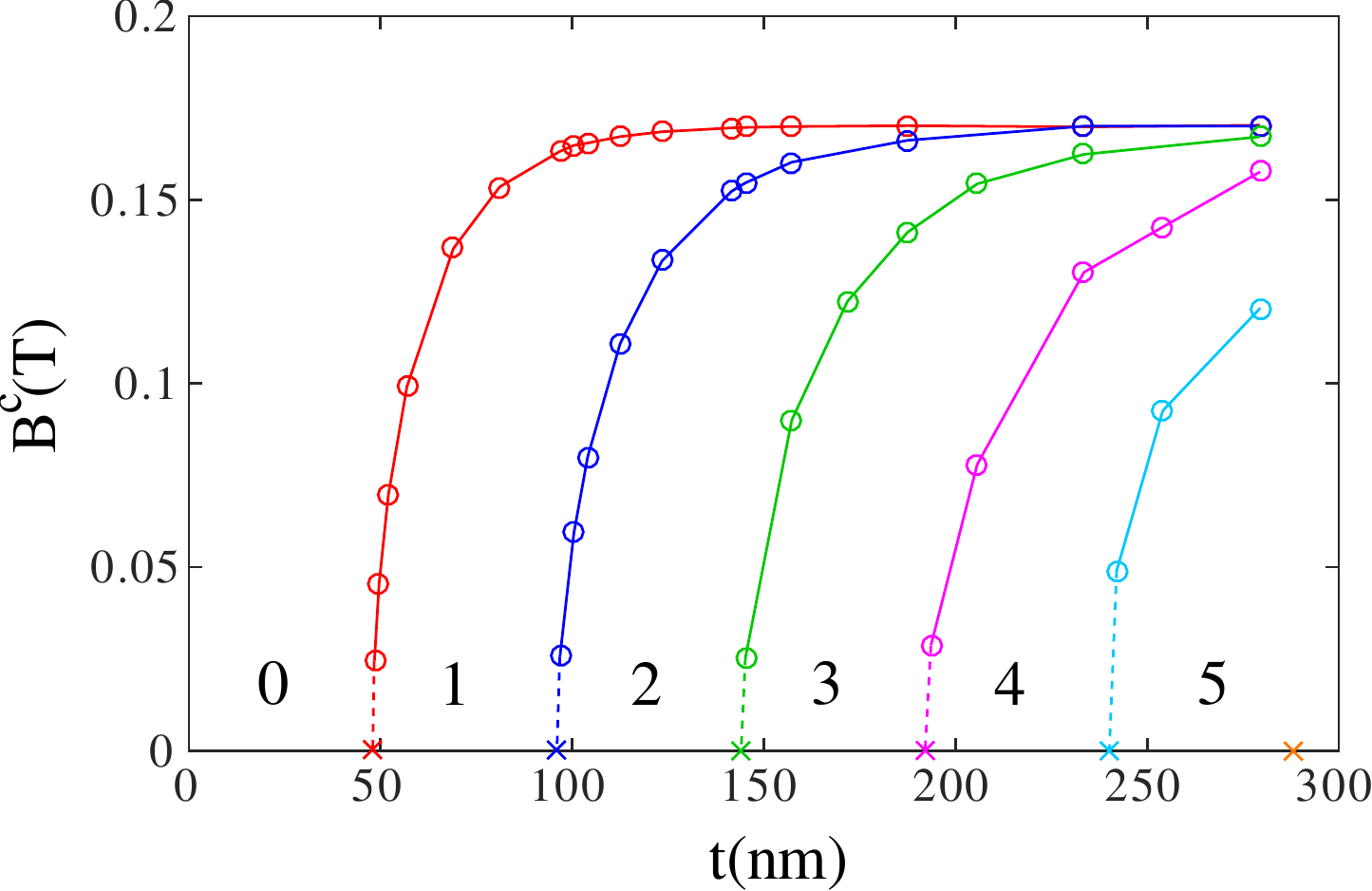}
\caption{\label{fig1} (Color online) Phase diagram of Cr$_{1/3}$NbS$_2$ as a function of thickness $t$ and applied external magnetic field $B$. The numbers correspond to topological index (the number of solitons) in each phase. The circles show the phase boundary obtained numerically, while crosses on the t-axes correspond to $t=nL_0$, where $n$ is an integer and $L_0=48$nm is the spiral period in zero field.}
\end{figure}

In the main text we stated that each transition between topological sectors observed upon increasing the in-plane magnetic field $B$ leads to a progressively better average alignment of the spin in the field direction. Since such a better alignment results in a lower measured resistance, this explains why the transitions between different topological are accompanied by the jumps in which the resistance value is systematically lowered. Here we use the theoretical model to calculate the evolution of the magnetization upon increasing the magnetic field, to show that the magnetization does indeed increase in "jumps" in correspondence of each transitions from one topological sector to another one with a smaller number of magnetic solitons.

Fig.\ref{fig1} presents the numerical phase diagram of Cr$_{1/3}$NbS$_2$ as a function of thickness $t$ and of applied external magnetic field $B$.
For samples of thickness $nL_0<t<(n+1)L_0$ ($n$ is a non-negative integer, $L_0=48$nm is the spiral period) there are $n$ phase transitions between topological sectors. As expected, all phase boundaries approach the bulk critical field $B\approx 0.17T$ when the thickness is increased. Starting from thicknesses about $t\approx 140$nm, phases with a small number of solitons are stabilized only in a very narrow interval of magnetic field $B$ and  will be difficult to detect experimentally (that is why in the main text we focused our quantitative analysis on the last transition, occurring at $B_c$).

\begin{figure}[h!]
\includegraphics[width=8cm]{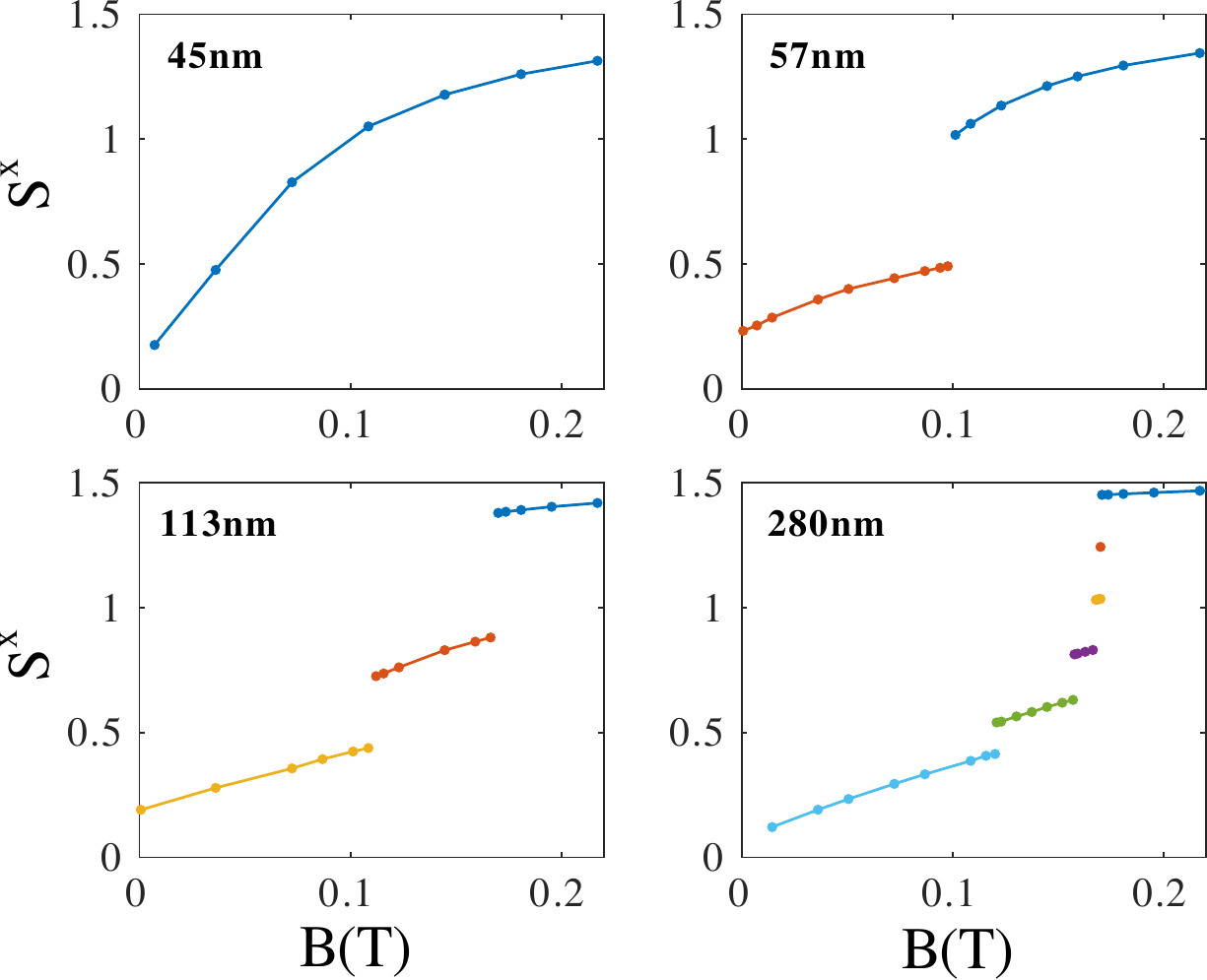}
\caption{\label{fig2} (Color online) Average magnetization as a function of the applied magnetic field for different thicknesses. }
\end{figure}

The multiple transitions between topological sectors observed experimentally in the magneto-resistance can be detected numerically by looking at the average magnetization:
\begin{equation}
S^x=\frac{1}{N}\sum_{i=1}^N S^x_i
\end{equation}
as a function of the magnetic field. Fig.\ref{fig2} presents these results for four different chain lengths. They agree qualitatively with the experimental results. As explained above, the jumps in the average magnetization are expected to correspond to the jumps in the magneto-resistance. 

\begin{figure*}[t!]
\includegraphics[width=\textwidth]{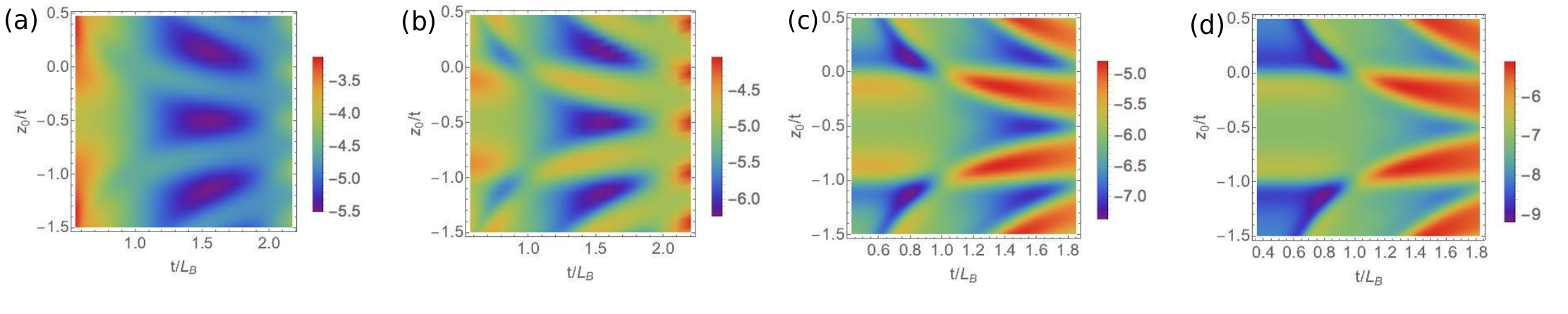}
\caption{\label{fig3} (Color online) The 2D density plot of energy (in units of $J$) as function of $t/L_B$ and $z_0/t$ for the case of $t= 1.6 L_0$ with different external field, (a) $h = 0.2$, (b) $h=0.4$, (c) $h=0.6$ and (d) $h=0.8$. The integer part of $t/L_B$ indicates the number of solitons in the film. The figures show the transition from the energetically favored one-soliton structure to zero-soliton structure, as the magnetic field is increased.  The exact transition point  for the case of  $t= 1.6 L_0$ is found to be  at $h=0.544$, which corresponds to $B=0.15$ T in the numerical calculation.  Note that in the figures, for the same value of $t/L_B$, there are two or three minimum energy points, which should correspond to the same spin configuration, since $z_0/t$ is related with only the phase shift. }
\end{figure*}

\section{Analytical calculation}
The above numerical results can be verified through an analytical calculation based on Hamiltonian (\ref{eq:H}) by reformulating it in the continuous limit \cite{Togawa2012,Togawa2013,Chapman2014,Bornstein2015}. Due to strong anisotropy in the $z$ direction, the spins are confined to be in the easy plane ($xy$ plane), and their configuration can be described by the azimuthal angle $\Phi(z)$. The total energy, therefore, is expressed as a functional of $\Phi(z)$,
\be
{\cal E}= J \int_0^t dz \Big(  \frac{1}{2} (\partial_z \Phi)^2 + q_0  (\partial_z \Phi) - h \cos\Phi \Big) \label{energy}
\ee
where $t$ is the film thickness, $q_0 = D/J$ is the wave vector of helix structure without external magnetic field, and $h = \mu_B B/J$ is the normalized magnitude of external field. Through this energy functional, $\Phi(z)$ should satisfy the following Sine-Gordon equation: 
\be
\frac{d^2 \Phi}{d z^2} = - h \sin \Phi, 
\ee
and its general solution is expressed as 
\be
\Phi = \pi - 2 \phi \Big(\frac{\sqrt{h} }{k} (z+z_0), k\Big) \label{Psi},
\ee
where $\phi(u, k)$ is the Jacobi amplitude with modulus $k$. This solution is parametrized by the two quantities $k$ and $z_0$, which should be determined by minimization of total energy with the appropriate boundary condition. Modulus $k$ determines the period of the soliton structure $L_B$ through 
\be
L_B= 2k K(k)/\sqrt{h},
\ee
where $K(k)$ is the complete  elliptical integral of the first kind, $z_0$ determines the phase shift of the structure. Once the two parameters are specified, the spin configuration is determined. 
For a bulk system, the parameter $z_0$ does not affect the energy and it has already been found that, to minimize the energy, $k$ should satisfy the condition:  
$4 \sqrt{h} E(k) - \pi k q_0=0$,
where $E(k)$ is the complete  elliptical integral of the second kind.  This condition gives rise to the critical field for bulk system, $h_c= \pi^2 /16 \approx 0.6185$, corresponding to the value of $B^c =0.17 $T obtained by numerical calculation. 

For a finite system, there is no particular reason to assume any preference in the $xy$ plane on the boundary, and therefore a free boundary condition is expected.  The energy functional can be explicitly written down as a function of thickness $t$, parameter $k$ and $z_0$
\be
{\cal E}/J = \Big(  2 \frac{\sqrt{h}}{ k} E(\phi_z, k) - 2 q_0 \phi_z + {\cal E}_h \Big) \Big|_{z=z_0}^{z=t+z_0}, \label{eq:energy}
\ee
with 
\be
{\cal E}_h = \frac{\sqrt{h}}{k} \Big( 2E(\phi_z, k) + (k^2 -2) F(\phi_z, k) \Big) , 
\ee
where we simply write $\phi_z \equiv \phi(\sqrt{h} z/k, k)$. $F(\phi, k)$ and $E(\phi, k)$ are the incomplete elliptic integral of the first and second kind, respectively.  The conditions on $z_0$ to minimize energy (\ref{eq:energy}) are found to be as: $z_0 = -t/2$ or $z_0 = (L_B -t)/2$. Even though the condition on parameter $k$ cannot be written down explicitly, energy functional form (\ref{eq:energy}) enables us  much more easily  to determine the numerical value of $k$ that minimizes the total energy (\ref{eq:H}). 
We will not repeat the results calculated based on (\ref{eq:energy}), that are in perfect agreement with numerical calculation. We just plot the energy as a function of $t/L_B$ and $z_0/t$  for different values of external field and vividly demonstrate the topological transition from one kind of soliton structure to another, as in Fig.~\ref{fig3}.

\end{document}